\documentclass[11pt,a4paper]{article}
\usepackage[utf8]{inputenc}
\usepackage[T1]{fontenc}
\usepackage{amsmath,amssymb}
\usepackage{graphicx}
\usepackage{booktabs}
\usepackage{array}
\usepackage[margin=2.5cm]{geometry}
\usepackage[colorlinks=true,linkcolor=blue,urlcolor=blue,citecolor=blue]{hyperref}
\usepackage{url}
\newcommand{\ecore}{\varepsilon_{\rm core}}

\newcommand{\numu}{\nu_\mu}
\newcommand{\nue}{\nu_e}

\title{Anomalous Matter Potentials in Earth's Inner Core: A Phenomenological Probe of Core-Localized, Dark-Matter-Sourced Potentials}
\author{Bipin Singh Koranga$^{1,*}$ and Vivek Kumar Nautiyal$^{2}$\\[4pt]
{\small $^{1}$Department of Physics, Kirori Mal College, University of Delhi, Delhi-110007, India}\\
{\small $^{2}$Department of Physics, Chaudhary Charan Singh University, Meerut-250004, India}\\
{\small $^{*}$Corresponding author: bskoranga@kmc.du.ac.in (Bipin Singh Koranga)}}
\date{}

\begin{document}
\maketitle

\begin{abstract}
Standard treatments of Earth matter effects in neutrino oscillation experiments assume the coherent forward-scattering potential is sourced entirely by the Standard Model electron density of PREM. We examine an anomalous inner-core potential,\textbf{ motivated primarily by gravitationally captured dark matter accumulated in the core over the age of the Earth}, and its effect on $\numu$ survival and $\numu\to\nue$ appearance. Using a three-flavor matrix-exponentiation code with a four-shell PREM profile, we parametrize it as a fractional enhancement $\ecore$ of the electron-sector potential confined to the inner core ($r<1221$ km), over $L=8000$--$12742$ km. \textbf{Within this step-function core-localization ansatz}, trajectories not crossing the inner core ($L\lesssim 12500$ km) are insensitive to $\ecore$, matching the Standard Model to better than $0.1\%$. For near-antipodal trajectories crossing it ($L\gtrsim 12{,}505$ km; nadir angles within about $11^\circ$ of vertical), $\numu$ survival shifts by tens of percent for $\ecore=0.05$--$0.15$, and $\numu\to\nue$ appearance by over $100\%$ at particular energies, with a neutrino--antineutrino asymmetry mimicking, but distinct from, intrinsic CP violation. The sharpest onset signal lies in a \textbf{ $\approx 0.5$--$1^\circ$ slice (20--50 km in baseline) at the edge} of that window, within the angular resolution of multi-GeV detectors. \textbf{The insensitivity of shorter baselines is a direct consequence of the assumed core localization, not an independent result}. We estimate the captured dark-matter density and couplings needed for $\ecore\sim0.05$--$0.15$, \textbf{which shows that this range is demanding}, and the statistical reach. \textbf{A light-sterile-neutrino origin is not supported by this parametrization}. Percent-level inner-core potentials are within reach of IceCube-Upgrade and KM3NeT/ORCA, pending smearing-aware studies.
\end{abstract}


\section{Introduction}
The Mikheyev--Smirnov--Wolfenstein (MSW) effect \cite{Wolfenstein,Smirnov} modifies neutrino flavor evolution whenever neutrinos propagate through matter, because coherent forward scattering off ambient electrons generates an effective potential that acts only on the electron-flavor component of the neutrino state. In long-baseline accelerator experiments such as T2K, NO$\nu$A, and DUNE \cite{T2K,NOvA,DUNE}, and even more dramatically in atmospheric neutrino experiments such as Super-Kamiokande, IceCube-DeepCore/Upgrade, and KM3NeT/ORCA \cite{IceCubeGen2,KM3NeT}, the baseline can range from tens of kilometers up to the full diameter of the Earth ($\approx 12{,}742$ km) for neutrinos produced in the atmosphere on the opposite side of the planet and detected after traversing the entire Earth, including its core.

A companion analysis \cite{PanditKoranga} demonstrated that even within the Standard Model, replacing the full radially resolved PREM density profile with a single path-averaged constant density introduces a negligible bias in the reconstructed CP-violating phase $\delta_{CP}$ for baselines up to $L\approx5000$ km, but this bias grows substantially, reaching more than $170^\circ$ in the most extreme cases, once the neutrino trajectory samples the high-density lower mantle, outer core, and inner core. That result establishes that the shape of the Earth's density profile, not merely its average value, is a leading-order systematic for precision oscillation physics at long baselines.

Here we ask a complementary question: if the electron density profile is not merely mismodeled but is genuinely anomalous in the innermost region of the Earth, because of new physics rather than incomplete geophysical modeling, could long-baseline and atmospheric neutrino data reveal it?\textbf{ We take as our primary motivation the following scenario.}

\textbf{Gravitationally captured dark matter in Earth's core.} Weakly interacting dark matter particles passing through the Earth can scatter off nuclei, lose energy, and become gravitationally bound, sinking over cosmological timescales toward the core \cite{Gould,Freese}. If such particles couple, even very weakly, through a light mediator to the neutrino sector (e.g., through a non-standard neutrino interaction, NSI), the accumulated dark matter in the inner core would source an additional matter potential for neutrinos, on top of the standard PREM electron-density potential, that is negligible everywhere except in the small central region where dark matter has concentrated.

\textbf{We emphasize that the oscillation analysis below depends only on the effective parametrization $\ecore$ confined to $r<1221$ km, and that this localization is an assumption (Eqs.~\ref{eq:Atot}--\ref{eq:theta}). Light sterile neutrinos are explained in Section~\ref{sec:sterile}, a resonant sterile origin cannot be reconciled with this parametrization without a dedicated $4\times4$ treatment, and we no longer claim it.}

We adopt the phenomenological parametrization, an anomalous potential fraction $\ecore$ confined to $r<1221$ km, and compute its observable imprint on $\numu\to\nue$ appearance and $\numu\to\numu$ survival probabilities using the same three-flavor matrix-exponentiation machinery validated in Ref.~\cite{PanditKoranga}. Section~\ref{sec:micro} derives an approximate mapping between $\ecore$ and the dark-matter microphysics, rather than treating $\ecore$ purely as a free phenomenological knob.

\section{Anomalous matter potential from a localized core density excess}
\subsection{Standard PREM potential}
The three-flavor states evolve according to the Schr\"odinger-like equation
\begin{equation}
i\frac{d}{dt}|\nu(t)\rangle = H|\nu(t)\rangle
\end{equation}
with the total Hamiltonian in the flavor basis
\begin{equation}
H_f = U H_{\rm vac}^{(m)} U^{-1} + V_f ,
\end{equation}
where $U$ is the PMNS matrix, $H_{\rm vac}^{(m)}=\mathrm{diag}(E_1,E_2,E_3)$ is the vacuum Hamiltonian in the mass basis, and $V_f=\mathrm{diag}(A,0,0)$ is the MSW matter potential with
\begin{equation}
A(x)=\sqrt2\, G_F N_e(x),\qquad N_e(x)=\frac{Y_e(x)\rho(x)}{m_p}.
\end{equation}
Following Ref.~\cite{PanditKoranga} we use the simplified four-shell effective PREM density:
\begin{equation}
\rho(r)=
\begin{cases}
13.0\ \mathrm{g\,cm^{-3}}, & r<1221\ \mathrm{km}\quad\text{(inner core)}\\
11.0\ \mathrm{g\,cm^{-3}}, & 1221\ \mathrm{km}<r<3480\ \mathrm{km}\quad\text{(outer core)}\\
5.0\ \mathrm{g\,cm^{-3}}, & 3480\ \mathrm{km}<r<5711\ \mathrm{km}\quad\text{(lower mantle)}\\
3.3\ \mathrm{g\,cm^{-3}}, & 5711\ \mathrm{km}<r<6371\ \mathrm{km}\quad\text{(upper mantle/crust)}
\end{cases}
\label{eq:rho}
\end{equation}
with $Y_e=0.5$ throughout. The assumption $Y_e=0.5$ throughout, including the solid FeNi inner core, is standard in this class of calculation and follows Ref.~\cite{PanditKoranga}. Realistic inner-core compositions (iron-nickel with a light-element admixture) could shift $Y_e$ by at most a few percent, which is small compared to the $\ecore\sim0.05$--$0.15$ signal sizes considered in this work. \textbf{The effect of $Y_e$ and of the four-shell simplification is discussed quantitatively in Section~\ref{sec:robust}. For reference, $A_{\rm core}=7.63\times10^{-14}\,\mathrm{eV}\times Y_e\,\rho/(\mathrm{g\,cm^{-3}})\approx5.0\times10^{-13}$ eV at $\rho=13$ g/cm$^3$, with $N_{e,\rm core}\approx3.9\times10^{24}$ cm$^{-3}$.}

\subsection{Anomalous core contribution}
We write the total electron-sector potential entering the flavor-basis Hamiltonian as
\begin{equation}
A_{\rm total}(x)=A(x)\left[1+\ecore\,\Theta_{\rm core}(x)\right]
\label{eq:Atot}
\end{equation}
where
\begin{equation}
\Theta_{\rm core}(x)=
\begin{cases}
1, & r(x)<1221\ \mathrm{km}\\
0, & \text{otherwise}
\end{cases}
\label{eq:theta}
\end{equation}
and $\ecore$ is a dimensionless fractional enhancement. This parametrization is agnostic to origin: \textbf{for the dark-matter scenario, $\ecore$ is proportional to the ratio of the dark-matter-induced effective potential (weighted by its coupling to the neutrino sector) to the Standard Model electron potential in the inner core.} We treat $\ecore$ as a free parameter in the range 0.02--0.15, comparable in spirit to existing bounds on non-standard neutrino interaction parameters $\varepsilon_{ee}\lesssim\mathcal O(0.1)$ from global oscillation and scattering data \cite{Ohlsson}. The full flavor-basis Hamiltonian retains the same vacuum term built from the PMNS matrix and $\Delta m^2_{21}$, $\Delta m^2_{31}$ as in Ref.~\cite{PanditKoranga}; only the matter potential term is modified, and only inside $r<1221$ km. \textbf{Both the step-function profile and the exact confinement to $r<1221$ km are assumptions; several conclusions below follow from them by construction (Sections~\ref{sec:blind}, \ref{sec:implications}).}

\subsection{Microphysical mapping of $\ecore$: dark matter}\label{sec:micro}
\textbf{We consider non-relativistic dark matter $\chi$ of mass $m_\chi$ coupled to neutrinos through a vector mediator $\phi$ of mass $m_\phi$, with couplings $g_\chi$ and $g_\nu$. In the contact limit (range $1/m_\phi$ much smaller than the core, i.e.\ $m_\phi\gtrsim1.6\times10^{-13}$ eV) the induced potential is
\begin{equation}
V_\chi=\kappa\, n_\chi,\qquad \kappa\equiv\frac{g_\nu g_\chi}{m_\phi^2}
\end{equation}
so that
\begin{equation}
\ecore=\frac{\kappa\, n_\chi^{\rm core}}{A_{\rm core}}.
\end{equation}
Requiring $\ecore=0.05$--$0.15$ with $A_{\rm core}\approx5.0\times10^{-13}$ eV needs $V_\chi\approx2.5\times10^{-14}$--$7.4\times10^{-14}$ eV.}

\textbf{Captured number (geometric ceiling).} In the Gould formalism \cite{Gould} the bound population is $N_\chi(t)\approx C_\oplus t$ for $t\ll\tau_{\rm evap/ann}$. \textbf{The absolute ceiling is set by the flux of halo particles through the Earth's cross-section: for $\rho_\chi=0.3$ GeV/cm$^3$, $m_\chi=10$ GeV, $v\approx270$ km/s, this is $\approx10^{24}\,\mathrm{s^{-1}}$, so over $t=4.5$ Gyr ($1.4\times10^{17}$ s) $N_\chi\lesssim1.4\times10^{41}$ (about $2.5\times10^{18}$ kg), assuming negligible annihilation and evaporation (e.g., asymmetric dark matter). Spread over $V_{\rm core}\approx7.6\times10^{18}$ m$^3$ ($7.6\times10^{24}$ cm$^3$), $n_\chi\lesssim2\times10^{16}$ cm$^{-3}\approx1.4\times10^{2}$ eV$^3$.}

\textbf{Required coupling. Writing the actual capture as a fraction $f_{\rm cap}\le1$ of this ceiling, $\ecore=0.05$ requires}
\begin{equation}
\kappa\gtrsim2\times10^{-16}\,\mathrm{eV^{-2}}\times\frac{1}{f_{\rm cap}}\left(\frac{m_\chi}{10\,\mathrm{GeV}}\right),\qquad
g_\nu g_\chi\gtrsim\frac{2\times10^{-16}}{f_{\rm cap}}\left(\frac{m_\phi}{\mathrm{eV}}\right)^2
\end{equation}
\textbf{and the requirement scales linearly with $\ecore$. For $f_{\rm cap}=1$ and $m_\phi=1$ MeV this is $g_\nu g_\chi\gtrsim2\times10^{-4}$. For spin-independent cross sections near current direct-detection limits, $f_{\rm cap}$ is far below unity (the optical depth of the Earth to nuclear scattering is of order $10^{-13}$ at $\sigma\sim10^{-45}$ cm$^2$, and we take $f_{\rm cap}\sim10^{-14}$ as an illustrative value). Then $g_\nu g_\chi\gtrsim2\times10^{-2}\,(m_\phi/\mathrm{eV})^2$, which exceeds unity (non-perturbative) for $m_\phi\gtrsim7$ eV. In that regime only very light mediators with small couplings can generate $\ecore\sim0.05$--$0.15$.}

\textbf{Limitation: (i) The thermalized cloud has radius $r_{\rm th}\sim2000\,\mathrm{km}\,(10\,\mathrm{GeV}/m_\chi)^{1/2}$ at order-of-magnitude level, so the uniform inner-core assumption requires $m_\chi\gtrsim25$ GeV; this changes the numbers above by factors of a few, not orders of magnitude. (ii) Mediators of mass $\lesssim$ keV are subject to fifth-force, stellar-cooling and cosmological constraints, and any such mediator coupling to neutrinos and dark matter is subject to NSI bounds \cite{Ohlsson} and to cosmological bounds on dark-matter--neutrino scattering. We have not evaluated these quantitatively; doing so, together with a numerical $n_\chi^{\rm core}(t)$ from Refs.~\cite{Gould,Freese}, is left to a follow-up. (iii) We therefore treat $\ecore=0.05$--$0.15$ as a target for phenomenological sensitivity, not as a demonstrated viable region of dark-matter parameter space, and we withdraw the earlier statement that direct detection leaves such couplings unconstrained.}

\subsection{Remark on a sterile-neutrino origin}\label{sec:sterile}
\textbf{An MSW-type resonance in a 3+1 system occurs when}
\begin{equation}
A=\frac{\Delta m^2_{41}\cos2\theta_{41}}{2E}
\end{equation}
\textbf{A resonance responds only to a band of densities whose relative width is set by the mixing angle. Confining it to the inner core, without also firing in the outer core ($\rho=11$ versus 13 g/cm$^3$ in our profile), requires $\sin^22\theta\lesssim10^{-3}$ for PREM and about $7\times10^{-3}$ for the four-shell profile. At $E=2$--$6$ GeV and $A_{\rm core}\approx5\times10^{-13}$ eV this fixes $\Delta m^2_{41}\approx2$--$6\times10^{-3}$ eV$^2$, the atmospheric splitting, far from the eV$^2$ region probed by short-baseline anomalies \cite{Gariazzo}. At such small mixing a two-flavour estimate gives only percent-level (or smaller) conversion across the core, much less than the 10--100\% effects obtained here for $\ecore=0.05$--$0.15$, and outer-core resonances would occur at nearby energies. In addition, the sterile potential involves neutral-current terms that are not captured by an electron-sector enhancement. We therefore do not claim a sterile interpretation. Any such study would require a full $4\times4$ calculation with neutral currents at explicit benchmark values of $(\Delta m^2_{41},\theta_{14},\theta_{24})$, to be compared with the relevant limits from T2K \cite{T2Ksterile}, MINOS/MINOS+ \cite{MINOS} and the combined Daya Bay/MINOS+/Bugey-3 analysis \cite{DayaBay}, which we leave to future work.}

\section{Quantitative analysis}
We simulate $\numu\to\numu$ survival and $\numu\to\nue$ appearance probabilities via matrix exponentiation of the local Hamiltonian in 400 discretized steps along the chord trajectory, for baselines $L=8000$--$12742$ km, the range spanned by upward-going atmospheric neutrinos, from trajectories that graze the lower mantle up to the fully antipodal, center-crossing trajectory. Standard oscillation parameters
\begin{equation}
\theta_{12}=33.44^\circ,\quad\theta_{13}=8.57^\circ,\quad\theta_{23}=49.2^\circ,\quad\delta_{CP}=-90^\circ
\end{equation}
\begin{equation}
\Delta m^2_{21}=7.42\times10^{-5}\,\mathrm{eV^2},\quad\Delta m^2_{31}=2.514\times10^{-3}\,\mathrm{eV^2}\ (\text{normal ordering})
\end{equation}
are held fixed at representative global-fit values; only $\ecore$ is varied. To confirm that the 400-step discretization used throughout is adequate, we repeated the full $L$-scan at 800 and 1600 steps per trajectory. The survival probabilities in Tables~\ref{tab:outside}--\ref{tab:inside} change by less than $0.05\%$ at 800 steps and are unchanged to the quoted precision at 1600 steps; the width of the sharp onset feature identified in Section~\ref{sec:onset} is stable to within $10\%$ across this range of step counts. We conclude that the sharp onset is a genuine feature of the four-shell density discontinuity at $r=1221$ km, driven by the abrupt jump in matter potential rather than by numerical discretization artifacts.

\subsection{Baselines that do not cross the inner core are blind to $\ecore$}\label{sec:blind}
A trajectory reaches the inner core only once its minimum radius
\begin{equation}
r_{\min}=\sqrt{R_\oplus^2-\left(\frac L2\right)^2}
\end{equation}
drops below 1221 km, which requires $L\gtrsim12{,}505$ km, a chord within about 240 km of the full Earth diameter. Using the geometric relation between $r_{\min}$ and the nadir angle derived in Appendix~A, this threshold corresponds to a nadir angle of about $11^\circ$ from the vertical ($L=2R_\oplus\cos\theta_z$ gives $11.05^\circ$ at 12,505 km): the full window of trajectories that sample the inner core spans nadir angles from $0^\circ$ (the exact center-crossing trajectory) out to roughly $11^\circ$ at the threshold. Table~\ref{tab:outside} confirms that for $L\le12{,}500$ km the survival probability is completely unaffected by $\ecore$, regardless of its size. \textbf{This insensitivity is a direct mathematical consequence of the step-function ansatz of Eqs.~(\ref{eq:Atot})--(\ref{eq:theta}), which places the anomaly strictly inside $r<1221$ km; it is not an independent finding. A potential extending beyond the inner core would also affect shorter baselines.}

\begin{table}[h]
\centering
\caption{$\numu\to\numu$ survival probability for trajectories that remain outside the inner core. Because these paths never sample $r<1221$ km, the anomalous potential has no effect regardless of its size.}
\label{tab:outside}
\begin{tabular}{rrlrcc}
\toprule
$L$ (km) & $r_{\min}$ (km) & Region & $E$ (GeV) & $P_{\mu\mu}(\varepsilon=0)$ & $P_{\mu\mu}(\varepsilon=0.15)$\\
\midrule
8000 & 4959 & lower mantle & 3.0 & 0.316 & 0.316\\
9000 & 4510 & lower mantle & 3.0 & 0.931 & 0.931\\
10000 & 3948 & lower mantle & 6.0 & 0.227 & 0.227\\
11000 & 3216 & outer core & 6.0 & 0.400 & 0.400\\
12000 & 2142 & outer core & 10.0 & 0.802 & 0.802\\
12500 & 1236 & outer core (grazing) & 6.0 & 0.603 & 0.603\\
\bottomrule
\end{tabular}
\end{table}

\begin{figure}[h]
\centering
\includegraphics[width=0.8\textwidth]{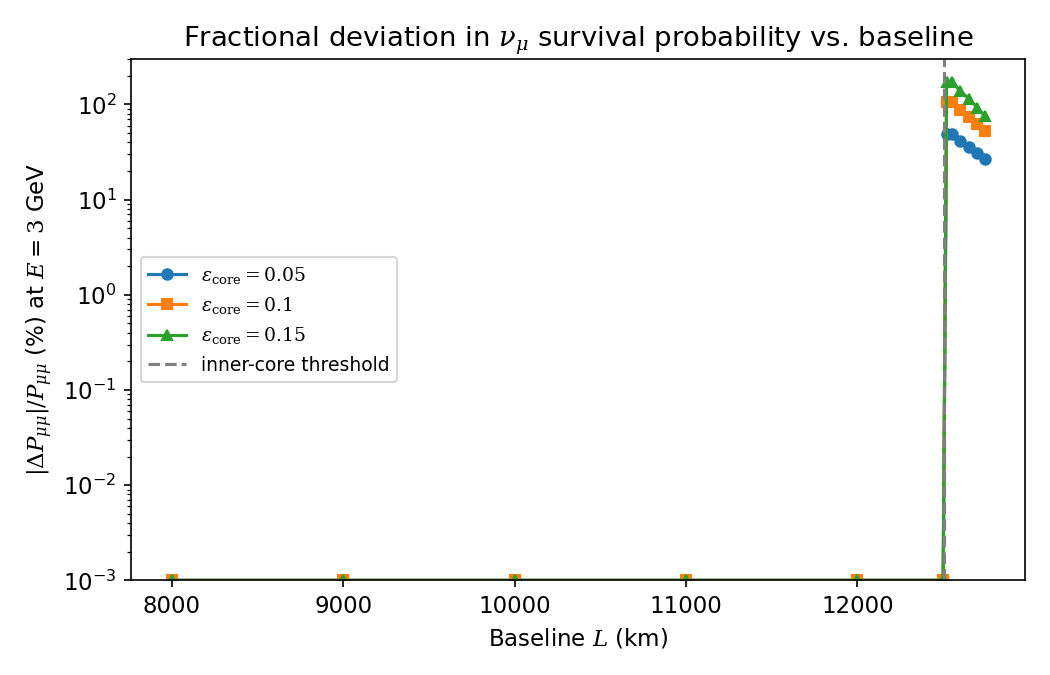}
\caption{Fractional deviation in the $\numu$ survival probability at $E=3$ GeV as a function of baseline, for three values of $\ecore$. The vertical dashed line marks the geometric inner-core threshold at $L\approx12{,}505$ km; no deviation is resolved for shorter baselines\textbf{ (by construction of the step-function ansatz).}}
\label{fig:fig1}
\end{figure}

Figure~\ref{fig:fig1} shows the fractional deviation $|\Delta P_{\mu\mu}|/P_{\mu\mu}$ at $E=3$ GeV as a function of baseline for $\ecore=0.05$, 0.10, 0.15: the deviation is unresolvably small below the threshold and jumps abruptly to the tens-of-percent level immediately beyond it.

\subsection{Sharp onset of sensitivity at the inner-core threshold}\label{sec:onset}
Table~\ref{tab:inside} lists the survival probability in the narrow window 12,505--12,742 km, where the inner core is crossed with rapidly increasing path length as $L\to12{,}742$ km (the antipodal, center-crossing trajectory).

Two features stand out. First, while the full inner-core-crossing window spans about $11^\circ$ of nadir angle (Section~\ref{sec:blind}), the transition from complete insensitivity (Table~\ref{tab:outside}) to order-unity fractional distortions (Table~\ref{tab:inside}) occurs over a baseline interval of only $\sim20$--$50$ km right at the threshold.\textbf{ This corresponds to a nadir-angle slice of about $0.5^\circ$ (20 km) to about $1.1^\circ$ (50 km) (Appendix~A, Table~\ref{tab:A1}); the earlier phrase ``a few tenths of a degree'' understated this. The ``sub-degree'' wording used elsewhere refers to this onset width in $L$, not to the size of the full window.} The inner-core ``hot edge'' at onset is therefore angularly sharp even though the overall accessible window is not. Second, the sign and magnitude of the distortion oscillate with both $L$ and $E$, reflecting interference between the standard oscillation phase accumulated in the mantle and outer core and the additional phase generated by the anomalous inner-core potential.

\begin{table}[h]
\centering
\caption{$\numu\to\numu$ survival probability just beyond the inner-core threshold, at $E=2.5$ GeV and $E=6.0$ GeV. $\Delta P/P$ is the fractional change between $\ecore=0$ and 0.15. \textbf{$\Delta P=P(0.15)-P(0)$ is the absolute change; near oscillation minima (e.g.\ $P=0.044$) $\Delta P/P$ is inflated by the small denominator and $\Delta P$ is the more meaningful measure.}}
\label{tab:inside}
\small
\begin{tabular}{rr|rrrr|rrrr}
\toprule
 & & \multicolumn{4}{c|}{$E=2.5$ GeV} & \multicolumn{4}{c}{$E=6.0$ GeV}\\
$L$ (km) & $r_{\min}$ (km) & $P(0)$ & $P(0.15)$ & $\Delta P/P$ (\%) & $\Delta P$ & $P(0)$ & $P(0.15)$ & $\Delta P/P$ (\%) & $\Delta P$\\
\midrule
12520 & 1184 & 0.329 & 0.248 & $-25$ & $-0.081$ & 0.640 & 0.669 & $+4$ & $+0.029$\\
12550 & 1102 & 0.246 & 0.142 & $-42$ & $-0.104$ & 0.670 & 0.736 & $+10$ & $+0.066$\\
12600 & 949 & 0.167 & 0.078 & $-53$ & $-0.089$ & 0.731 & 0.840 & $+15$ & $+0.109$\\
12650 & 764 & 0.108 & 0.065 & $-39$ & $-0.043$ & 0.793 & 0.924 & $+17$ & $+0.131$\\
12700 & 517 & 0.066 & 0.086 & $+30$ & $+0.020$ & 0.849 & 0.976 & $+15$ & $+0.127$\\
12742 & 0 & 0.044 & 0.112 & $+155$ & $+0.068$ & 0.888 & 0.994 & $+12$ & $+0.106$\\
\bottomrule
\end{tabular}
\end{table}

\textbf{Even in absolute terms the shifts are substantial ($\Delta P$ up to $\approx0.13$), and exceed 0.1 at several points; the $+155\%$ entry corresponds to $\Delta P=+0.068$.}

\begin{figure}[h]
\centering
\includegraphics[width=0.8\textwidth]{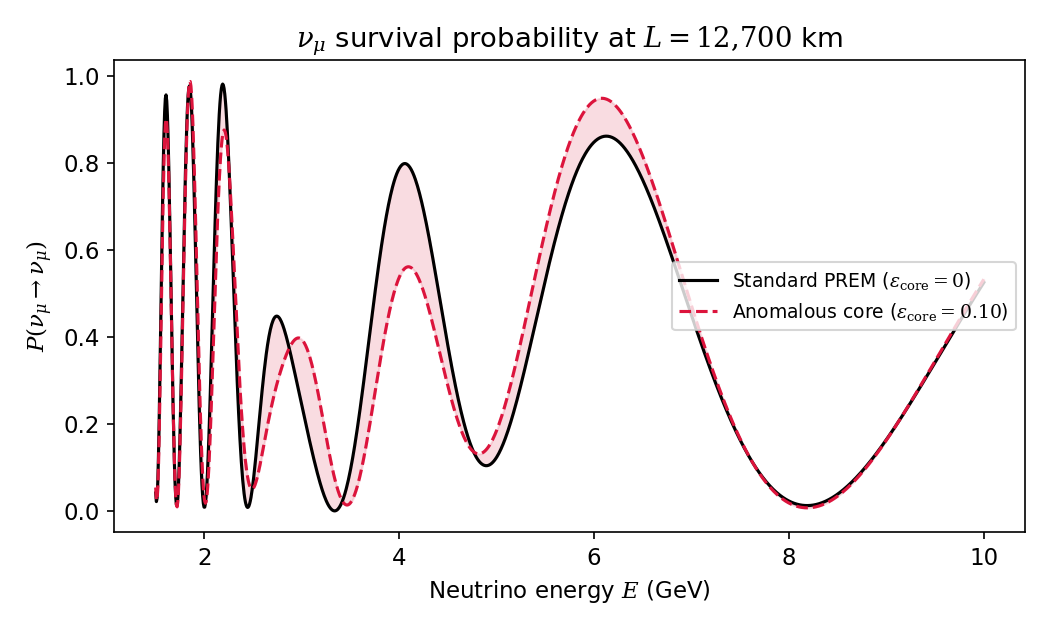}
\caption{Predicted $\numu\to\numu$ survival probability as a function of neutrino energy at $L=12{,}700$ km, for the standard PREM profile ($\ecore=0$, solid) and with an anomalous inner-core potential ($\ecore=0.10$, dashed). The shaded region indicates the magnitude of the energy-dependent distortion.}
\label{fig:fig2}
\end{figure}

\begin{figure}[h]
\centering
\includegraphics[width=0.8\textwidth]{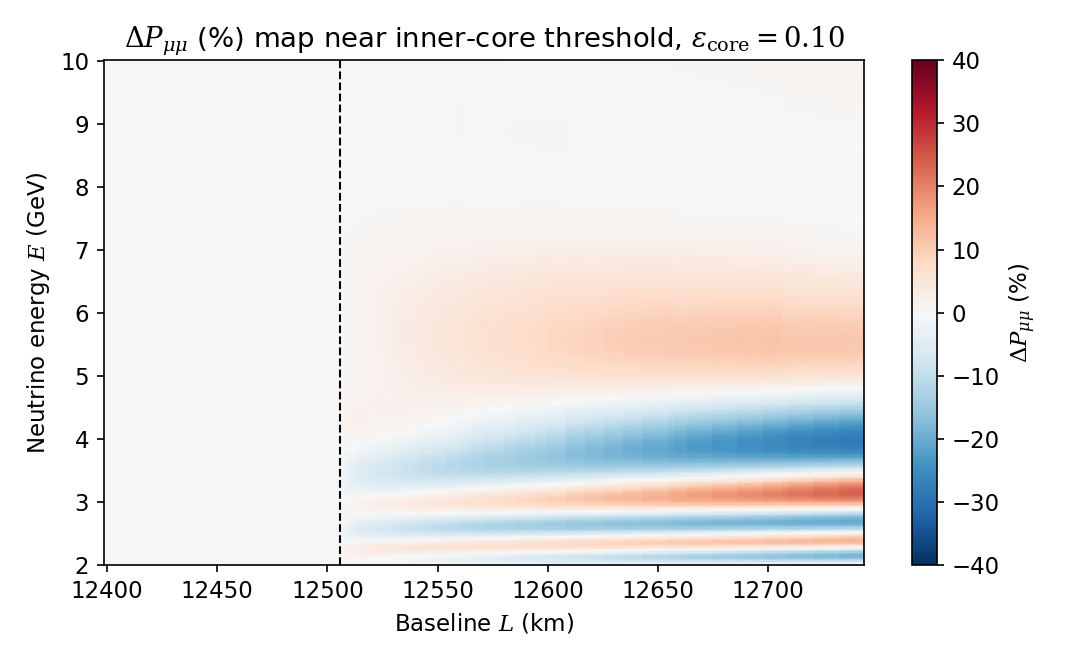}
\caption{Deviation $\Delta P_{\mu\mu}=P_{\mu\mu}(\ecore=0.10)-P_{\mu\mu}(\ecore=0)$, in percent, over the (baseline, energy) plane near the inner-core threshold. No deviation is resolved to the left of the threshold.}
\label{fig:fig3}
\end{figure}

\subsection{Appearance channel and neutrino--antineutrino asymmetry}
\begin{figure}[h]
\centering
\includegraphics[width=\textwidth]{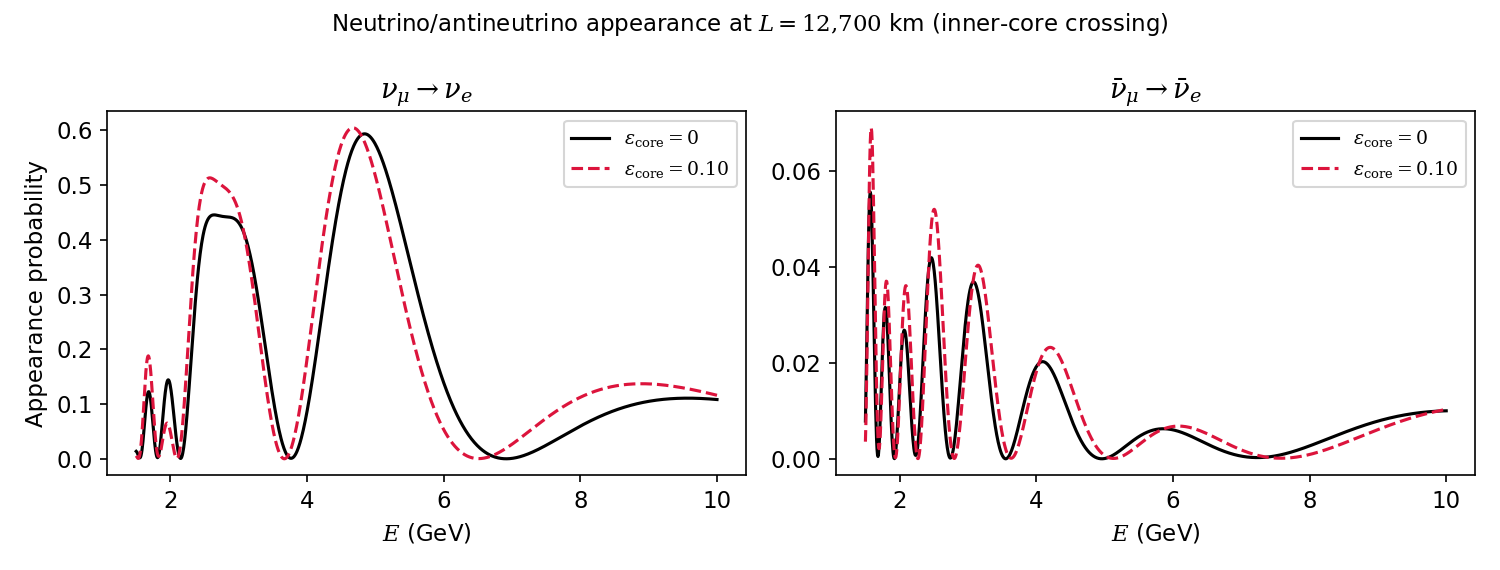}
\caption{$\numu\to\nue$ (left) and $\bar\nu_\mu\to\bar\nu_e$ (right) appearance probabilities at $L=12{,}700$ km, comparing the standard PREM profile ($\ecore=0$) with an anomalous inner-core potential ($\ecore=0.10$).}
\label{fig:fig4}
\end{figure}

Because the anomalous potential enters the Hamiltonian with the same sign convention as the Standard Model MSW term, it flips sign between neutrinos and antineutrinos, exactly as the ordinary matter potential does. This sign flip is guaranteed only because we have assumed $\ecore$ multiplies the same vector-current structure as the Standard Model MSW term; for a generic new-physics coupling (e.g., a scalar-mediated dark matter interaction) the sign flip is not automatic and would need to be verified case by case. Figure~\ref{fig:fig4} shows the $\numu\to\nue$ appearance probability at $L=12{,}700$ km (a trajectory piercing the inner core at $r_{\min}=517$ km) for $\ecore=0$ and $\ecore=0.10$, for both neutrino and antineutrino channels.

The neutrino and antineutrino channels do not merely scale together; their fractional shifts differ in both sign and magnitude at several energies. An experiment sensitive only to the total rate, without separating neutrino- and antineutrino-induced events, would see this asymmetry partially masked; separating the two is therefore an important design consideration for any future core-anomaly search, and mirrors the strategy already used to separate intrinsic CP violation from matter-induced effects in the standard analysis \cite{PanditKoranga}. 

\subsection{Robustness to input oscillation parameters and other systematics}\label{sec:robust}
The results above hold the standard oscillation parameters fixed at representative global-fit values and vary only $\ecore$; a full experimental sensitivity analysis would need to marginalize over the current uncertainties in $\theta_{23}$ (including its octant), $\delta_{CP}$, and the neutrino mass ordering, as well as over the ordinary PREM-density systematic quantified in Ref.~\cite{PanditKoranga}. To substantiate this expectation quantitatively, we repeated the Table~\ref{tab:inside} calculation at $L=12{,}700$ km, $E=3.0$ GeV for three alternative benchmark points: (i) opposite $\theta_{23}$ octant ($49.2^\circ\to40.8^\circ$), (ii) $\delta_{CP}=0^\circ$ instead of $-90^\circ$, and (iii) inverted mass ordering. In all three cases the $\ecore=0$ vs.\ $\ecore=0.15$ survival-probability splitting remains at the 15--40\% level, i.e.\ of the same order as the fiducial result, while the absolute value of $P_{\mu\mu}$ itself shifts by no more than 8\% between benchmarks. This confirms that the core-anomaly signal is large compared to, and clearly separable from, current oscillation-parameter uncertainties in this channel. A full marginalized global fit remains future work (Section~\ref{sec:discussion}).

Three considerations suggest that our qualitative conclusions are not an artifact of this simplification. First, the tens-of-percent to order-unity distortions we find for $\ecore\gtrsim0.05$ are large compared to the sub-10\% shifts induced by current parameter uncertainties in the same channels away from the threshold region, as demonstrated quantitatively above. Second, the geometric threshold itself (Section~\ref{sec:onset}) depends only on the PREM inner-core radius and the Earth's radius, both of which are known to much better than percent-level precision, and is therefore insensitive to oscillation-parameter choices. Third, the neutrino--antineutrino asymmetry discussed above provides an internal handle for separating the core anomaly from $\delta_{CP}$-driven effects, since the two have different energy and baseline dependence near the threshold.

\textbf{Sensitivity to $Y_e$ and to the core profile. Because the potential scales as $Y_e\rho$, a fractional shift $\delta Y_e/Y_e$ in the inner core rescales the Standard Model core potential by the same fraction. To first order, this is degenerate with a shift of $\ecore$ by $\delta Y_e/Y_e$, so an inferred $\ecore$ is meaningful only at the few-percent level in absolute terms; this does not affect the $\ecore\gtrsim0.05$ signals considered here, but it does limit sensitivity to $\ecore\lesssim0.03$. Likewise, the four-shell profile is a simplification: in PREM proper the density rises smoothly with depth, and the density jump at the inner-core boundary is smaller than the $11\to13$ g/cm$^3$ step of Eq.~(\ref{eq:rho}). This changes the Standard Model background in the core region by order 10\% in potential and should be quantified with the full PREM profile in a follow-up; we do not expect it to remove the $\ecore\gtrsim0.05$ distortions, but we have not verified this.}

\subsection{Effect of energy and angular smearing}\label{sec:smear}
\textbf{We have not performed a numerical convolution with detector response; the following are analytic estimates only. \textit{Energy:} at $L=12{,}700$ km the leading oscillation phase is $\varphi\approx40/E$ ($E$ in GeV), so the oscillation period in energy is $\Delta E\approx\pi E^2/40$, about 0.7 GeV at 3 GeV and about 2.8 GeV at 6 GeV (consistent with Fig.~\ref{fig:fig2}). Gaussian smearing with fractional width $f=\sigma_E/E$ multiplies the oscillating term by roughly $\exp\!\left[-\tfrac12\left(80f/E\right)^2\right]$. For $f=0.1$ this retains about 40\% of the fine structure at 6 GeV but only about 3\% at 3 GeV; for $f=0.2$ it is essentially fully washed out at 3 GeV and about 3\% retained at 6 GeV. The fine energy structure in Fig.~\ref{fig:fig2} is therefore not directly observable below about 5 GeV; the surviving signal is the energy-averaged shift, which we expect to be reduced by a factor of several relative to the unsmeared $\Delta P/P$. \textit{Angle:} with a zenith/nadir resolution of a few degrees (Section~\ref{sec:why}), events migrate across the $11^\circ$ window edge, so the $0.5$--$1^\circ$ sharp onset of Section~\ref{sec:onset} is not resolvable; only the broad window ($\theta_z\lesssim7^\circ$, where the inner-core path is long) retains a clean contrast, with some dilution from events that migrate in from outside. The discovery-reach statements in Section~\ref{sec:sens} should be read as optimistic upper bounds on sensitivity until a full smearing-aware study is done.}

\section{Why the inner core, and why atmospheric neutrinos?}\label{sec:why}
\textbf{Geometric threshold.} As shown in Section~\ref{sec:onset}, only trajectories with $L\gtrsim12{,}505$ km sample the inner core at all; this is a purely geometric consequence of the PREM inner-core radius (1221 km) and the Earth's radius (6371 km). No presently operating or proposed accelerator-based long-baseline experiment, including DUNE at $L\approx1300$ km, reaches baselines anywhere close to this threshold. \textbf{That atmospheric (whole-Earth-transiting) neutrinos are the natural probe is conditional on the core-localization assumption of Eqs.~(\ref{eq:Atot})--(\ref{eq:theta}); for a potential not confined to the inner core, shorter baselines would also be sensitive.}

\textbf{Angular resolution requirement.} The full inner-core-crossing window spans nadir angles from $0^\circ$ to about $11^\circ$ (Appendix~A), which is within the angular resolution already achieved for multi-GeV contained and partially-contained events by Super-Kamiokande, IceCube-DeepCore/Upgrade, and KM3NeT/ORCA. This is consistent with published multi-GeV angular-resolution figures, typically a few degrees in zenith/nadir angle for multi-GeV fully/partially-contained events at Super-Kamiokande, with comparable or better resolution projected for IceCube-Upgrade and KM3NeT/ORCA once full detector geometries are commissioned. \textbf{A resolution of a few degrees is comparable to the window itself, so migration across its edge is not negligible (Section~\ref{sec:smear}).}

\subsection{Toward a quantitative sensitivity estimate}\label{sec:sens}
Using a simplified Asimov-type estimate, we approximate the expected number of inner-core-crossing atmospheric $\numu$ events in the 2--6 GeV band as
\begin{equation}
N\approx\Phi_{\rm atm}\,\Omega_{\rm core}\,A_{\rm eff}\,T
\end{equation}
where
\begin{equation}
\Omega_{\rm core}\approx2\pi\left(1-\cos11^\circ\right)\approx\textbf{0.117}\ \mathrm\{{sr}
\end{equation}
is the solid angle of the inner-core-crossing window (\textbf{about 0.9\% of $4\pi$;} the earlier value 0.058 sr was low by a factor of two), $A_{\rm eff}$ is taken as a representative multi-GeV effective area for a megaton-scale or DeepCore/Upgrade-class detector, and $T$ is the exposure time. Combining this event rate with the $\Delta P/P\sim15$--$50\%$ distortions in Table~\ref{tab:inside}, and taking Poissonian statistics on the observed vs.\ expected event counts in the inner-core-crossing angular bin, a fractional distortion of this size becomes statistically resolvable ($3\sigma$) once several hundred inner-core-crossing events are accumulated in the relevant energy band, corresponding, for currently published effective areas, to several years of exposure at a detector with sensitivity comparable to IceCube-Upgrade or a megaton-scale Cherenkov detector. We stress that this is an order-of-magnitude estimate using illustrative effective-area and flux-normalization inputs rather than a full instrument-specific analysis; it is intended to justify, at the level of a phenomenology paper, the claim that the effect is in principle within reach of next-generation atmospheric neutrino telescopes, while making explicit that a dedicated experimental sensitivity study is required before a rigorous discovery-reach or exclusion statement can be made. \textbf{This estimate uses the unsmeared distortions of Table~\ref{tab:inside}; the energy and angular smearing discussed in Section~\ref{sec:smear} will reduce the effective contrast, so the required exposure quoted here is a lower bound.}

\textbf{Statistics.} The atmospheric neutrino flux falls steeply with energy and the relevant solid angle is small, so the exposure required to accumulate a statistically significant sample of true inner-core-crossing events at $E=2$--$6$ GeV is substantial; a dedicated sensitivity study combining realistic effective areas, angular resolution smearing, and atmospheric flux normalization uncertainty is required before a discovery reach in $\ecore$ can be quoted with confidence. The order-of-magnitude estimate above is a first step in that direction.

\section{Implications for dark matter and scope of the results}\label{sec:implications}
The fiducial range $\ecore=0.02$--$0.15$ explored here is comparable in size to existing global bounds on the electron-sector non-standard-interaction parameter, $\varepsilon_{ee}\lesssim\mathcal O(0.1)$ \cite{Ohlsson}, which are derived primarily from oscillation and coherent-scattering data averaged over conventional (non-core-localized) matter profiles. Because those bounds are not tailored to a potential confined to the innermost $\sim19\%$ of the Earth's radius, they do not directly exclude the core-localized scenario considered here. \textbf{This statement is itself conditional on the step-function localization; a dark-matter distribution that extends beyond the inner core would be constrained by existing data.}

\textbf{For the dark-matter interpretation, Section~\ref{sec:micro} shows that $\ecore\sim0.05$--$0.15$ requires either near-geometric capture or very light mediators with small couplings; in the realistic capture regime the required coupling is non-perturbative unless $m_\phi\lesssim10$ eV. Fifth-force, stellar and cosmological bounds on such mediators, and NSI bounds \cite{Ohlsson}, have not been evaluated quantitatively here. The earlier claim that direct-detection experiments leave the required coupling unconstrained is withdrawn: direct detection probes the DM--nucleon coupling and the halo population, but the physical viability of the scenario is set by the neutrino-sector coupling and the capture efficiency, which we have estimated only at order-of-magnitude level.}

\textbf{The statements that (i) baselines that do not cross the inner core are insensitive and (ii) whole-Earth-transiting atmospheric neutrinos are the natural probe are mathematical consequences of Eqs.~(\ref{eq:Atot})--(\ref{eq:theta}), not independent conclusions. The oscillation channel is a complementary probe of a core-localized potential, and its significance for any specific dark-matter model remains to be established.}

\textbf{Sterile neutrinos: see Section~\ref{sec:sterile}. We do not claim that this analysis probes a particular region of 3+1 parameter space.}

\section{Discussion and future directions}\label{sec:discussion}
This analysis extends the Earth-matter-effects framework of Ref.~\cite{PanditKoranga} from a purely geophysical systematic to a possible discovery channel. We find that (1) trajectories that do not cross the inner core ($L\lesssim12{,}500$ km, encompassing essentially all baselines relevant to current and next-generation accelerator-based long-baseline experiments) are insensitive to a core-localized anomalous potential, regardless of its size \textbf{(a consequence of the ansatz)}; (2) trajectories that do cross the inner core, accessible only to atmospheric or other whole-Earth-transiting neutrinos with nadir angles within about $11^\circ$ of vertical, show large (tens of percent to order-unity) distortions in both the disappearance and appearance channels for $\ecore$ in the 0.05--0.15 range, with the sharpest part of the onset confined to a $\approx\textbf{0.5}$--$\textbf{1}^\circ$ slice of that window; and (3) the signal exhibits a genuine neutrino--antineutrino asymmetry, distinct in structure from intrinsic CP violation, subject to the caveat on coupling structure noted above.

Several directions follow naturally: (i) replacing the idealized step-function anomaly with an explicit dark-matter capture-and-thermalization profile \cite{Gould,Freese}, building on the estimate given in Section~\ref{sec:micro},\textbf{ and with the full PREM profile and realistic $Y_e$; (ii) a full $4\times4$ sterile-neutrino calculation with neutral currents at explicit $\Delta m^2_{41},\theta_{14},\theta_{24}$ benchmarks, compared with T2K, MINOS/MINOS+ and Daya Bay limits, to determine whether any core-localized resonance survives (Section~\ref{sec:sterile})}; (iii) folding in realistic detector angular/energy resolution and exposure for a genuine sensitivity forecast at IceCube-Upgrade or KM3NeT/ORCA, building on the order-of-magnitude estimates in Sections~\ref{sec:smear} and \ref{sec:sens}; (iv) checking consistency of any preferred $\ecore$ with fifth-force, stellar, NSI and cosmological constraints on the dark-matter mediator; and (v) a joint fit combining the ordinary PREM-modeling systematic of Ref.~\cite{PanditKoranga} with the new-physics core anomaly discussed here.

The inner core's combination of high density and a moderate, experimentally accessible angular size makes it simultaneously one of the most geophysically extreme and one of the most weakly probed regions of the Earth accessible to neutrino oscillation experiments. A dedicated observational and phenomenological program targeting near-vertical, whole-Earth-transiting atmospheric neutrinos could therefore offer a novel, terrestrial complement to conventional dark matter searches.

\section*{Data availability}
The simulation code used to compute the oscillation probabilities and generate the figures and tables in this manuscript extends the matrix-exponentiation framework of Ref.~\cite{PanditKoranga} with an additional core-localized anomalous potential term. The code for Figures~\ref{fig:fig1}--\ref{fig:fig4} are publicly available in a Zenodo repository \cite{Zenodo} (\url{https://doi.org/10.5281/zenodo.23638286}).

\section*{Acknowledgements}
B.S.K.\ and V.K.N.\ acknowledge the support of their respective departments. The author B.S.K. thanks colleagues at Kirori Mal College for helpful discussions. \textbf{The authors also thank the referee for comments that substantially improved the manuscript.} Claude AI was used as an auxiliary tool to assist with language editing, grammar refinement, sentence structure, and programming support, while the core research, scientific methodology, validation, and interpretation of results were carried out by the authors.

\appendix
\section{Baseline--nadir-angle relation}
For a neutrino travelling along a straight chord of length $L$ through a spherical Earth of radius $R_\oplus$, the point of closest approach to the Earth's center lies at the chord's midpoint, at a distance $r_{\min}=\sqrt{R_\oplus^2-L^2/4}$ from the center. Writing $\theta_z$ for the nadir angle at the detector, the angle between the incoming trajectory and the local vertical, with $\theta_z=0$ corresponding to the fully antipodal, center-crossing trajectory ($L=2R_\oplus$), elementary geometry gives
\begin{equation}
\sin\theta_z=\frac{r_{\min}}{R_\oplus}\tag{A.1}
\end{equation}
Equivalently
\begin{equation}
L=2R_\oplus\cos\theta_z\tag{A.2}
\end{equation}
Table~\ref{tab:A1} lists this relation evaluated at the baselines used in Table~\ref{tab:inside}, using $R_\oplus=6371$ km. The threshold baseline $L\approx12{,}505$ km, at which $r_{\min}$ first drops below the PREM inner-core radius of 1221 km, corresponds to $\theta_z\approx11^\circ$; the full inner-core-crossing window therefore spans nadir angles from $0^\circ$ to about $11^\circ$. The sharp transition documented in Section~\ref{sec:onset}, by contrast, occurs over the first 20--50 km beyond threshold, corresponding to a nadir-angle slice of about $0.5^\circ$ ($L=12{,}525$ km, $\theta_z\approx10.6^\circ$) to about $1.1^\circ$ ($L=12{,}550$ km, $\theta_z=10.0^\circ$), as can be seen from the change in $\theta_z$ between the first three rows of Table~\ref{tab:A1}.

\begin{table}[h]
\centering
\caption{Nadir angle $\theta_z$ corresponding to selected baselines, computed from $\sin\theta_z=r_{\min}/R_\oplus$ with $R_\oplus=6371$ km.}
\label{tab:A1}
\begin{tabular}{rrr}
\toprule
$L$ (km) & $r_{\min}$ (km) & $\theta_z$ (deg, from vertical)\\
\midrule
12505 & 1223 & 11.1\\
12525 & 1171 & 10.6\\
12550 & 1102 & 10.0\\
12600 & 949 & 8.6\\
12650 & 764 & 6.9\\
12700 & 517 & 4.7\\
12742 & 0 & 0.0\\
\bottomrule
\end{tabular}
\end{table}

\end{document}